\documentclass[journal,onecolumn]{IEEEtran}

\ifCLASSINFOpdf
\usepackage[pdftex]{graphicx}
\else
\fi

\usepackage{amsthm}

\usepackage[cmex10]{amsmath}

\usepackage{algorithmic}

\usepackage{stfloats}

\usepackage{threeparttable}
\usepackage{newclude}
\usepackage{multirow}
\usepackage{amsfonts}
\usepackage{amsthm}
\usepackage{verbatim}
\usepackage{xcolor}
\usepackage{textcomp}		
\usepackage{epstopdf}
\usepackage{textcomp}
\usepackage[caption=false]{subfig}
\usepackage{tabularx}
\newcolumntype{P}[1]{>{\centering\arraybackslash}p{#1}}
\newcolumntype{M}[1]{>{\centering\arraybackslash}m{#1}}

\usepackage{color,soul}

\usepackage{siunitx}

\usepackage{cite} 

\begin{document}
	%

	\title{Electric Vehicle Charging Station Placement Method for Urban Areas}
	
	%
	%
	%
	\author{Qiushi~Cui,~\IEEEmembership{Member,~IEEE,}
            Yang~Weng,~\IEEEmembership{Member,~IEEE,}
		    and Chin-Woo~Tan,~\IEEEmembership{Member,~IEEE}
		\thanks{Qiushi Cui and Yang Weng are with the Department
			of Electrical and Computer Engineering, Arizona State University, Tempe, AZ, 85281 USA (e-mail: \mbox{qiushi.cui@asu.edu}; \mbox{yang.weng@@asu.edu}). Chin-Woo Tan is with Stanford University, Stanford, CA, 94305 USA (e-mail: \mbox{tancw@stanford.edu}).}
		


		
		}


\maketitle
	
\begin{abstract}
For accommodating more electric vehicles (EVs) to battle against fossil fuel emission, the problem of charging station placement is inevitable and could be costly if done improperly. Some researches consider a general setup, using conditions such as driving ranges for planning. However, most of the EV growths in the next decades will happen in the urban area, where driving ranges is not the biggest concern. For such a need, we consider several practical aspects of urban systems, such as voltage regulation cost and protection device upgrade resulting from the large integration of EVs. Notably, our diversified objective can reveal the trade-off between different factors in different cities worldwide. 
To understand the global optimum of large-scale analysis, we add constraint one-by-one to see how to preserve the problem convexity. Our sensitivity analysis before and after convexification shows that our approach is not only universally applicable but also has a small approximation error for prioritizing the most urgent constraint in a specific setup. 
Finally, numerical results demonstrate the trade-off, the relationship between different factors and the global objective, and the small approximation error. A unique observation in this study shows the importance of incorporating the protection device upgrade in urban system planning on charging stations. 


		
\end{abstract}
	


	%
\IEEEpeerreviewmaketitle
	
\section{Introduction}

\IEEEPARstart{U}{nder} the Paris agreement signed in $2016$, the model of a sustainable urban city -- Singapore, pledged to cut emissions intensity by $36\%$ below $2005$ levels by $2030$. To meet the commitment, emissions reduction worldwide in the transport sector is crucial, and large-scale electric vehicle (EV) adoption in the future is, therefore, utmost essential to Singapore and many other cities/countries. For example, Singapore took several important steps in this direction such as $1$) an announcement of a new Vehicular Emissions Scheme and 2) the launch of the electric vehicle car-sharing program, etc. However, one of the major barriers to successful adoption of EVs at a large scale is the limited number of available charging stations. Thus, it is important to properly deploy EV charging infrastructure to enhance the adoption of EVs efficiently. 




EV charging station placement has therefore been an active research area for intercity and urban infrastructure planning. In freeway charging infrastructure planning, \cite{ref:dong2016planning} tackles the EV charging station placement problem in a simple round freeway, whereas \cite{ref:zhang2017pev} proposes a capacitated-flow refueling location model to capture PEV charging demands in a more complicated meshed transport network. However, both papers share the similarity of considering the driving range in the freeway. In contrast, the driving range constraints are not prominent in the urban area charging infrastructure planning since the charging stations are easily accessible, therefore, researchers have considered various aspects dedicated for urban area charging station placement. For example, \cite{ref:xwang2017electric} manages to find the optimal way to recharge electric buses with long continuous service hours under two scenarios: with and without limited batteries. However, it is applicable only to public bus systems. \cite{ref:Rajabi2017optimal} considers urban traffic circulations and hourly load change of private EVs, but it ignores the geographical land and labor cost variation that are of high importance in urban areas. 


If zooming in on the specific techniques deployed and the realistic factors considered, the problem under study can be examined in various technical aspects. For example, \cite{ref:zheng2014electric} includes the annual cost of battery swapping, and \cite{ref:mehta2018smart} considers the vehicle-to-grid and grid-to-vehicle technology. Furthermore, researchers and engineers explore many realistic factors such as the investment and energy losses \cite{ref:yao2014multi}, quality of service \cite{ref:luo2017placement}, service radius \cite{ref:liu2013optimal}, etc. The work in \cite{ref:zhang2017placement} considers the EV integration impact on the grid from voltage and power perspectives. In fact, when the load profiles change, the electrical demand at particular points can exceed the rated value of the local T\&D infrastructure. A study in the U.S. has put the value of deferring network upgrade work at approximately \$$650$/kW for transmission and \$$1,050$/kW for distribution networks \cite{ref:dyke2010impact}. Besides the techniques in the previously mentioned papers, studies focusing on the infrastructure upgrade, therefore, seems necessary under large-scale EV integration. 




The aforementioned urban planning and technical issues are mainly formulated as optimization problems. Based on the nature of the equations involved, these optimization problems contain linear programming as well as the nonlinear programming problems \cite{ref:mehta2018smart}, \cite{ref:yao2014multi}. Based on the permissible values of the decision variables, integer programming, and real-valued programming usually, exist in the same EV charging station problem \cite{ref:lam2014electric}. Based on the number of objective functions, both single-objective \cite{ref:zhang2017pev} and multi-objective \cite{ref:yao2014multi,ref:wang2013traffic,ref:ruifeng2017multi} problems are proposed by researchers. Variously, the optimization problems are sometimes considered on a game theoretical framework in \cite{ref:luo2017placement}, \cite{ref:xiong2017optimal}. Solutions to these optimization problems include greedy algorithm \cite{ref:xwang2017electric}, \cite{ref:lam2014electric}, genetic algorithm \cite{ref:mehta2018smart}, interior point method \cite{ref:liu2013optimal}, gradient methods \cite{ref:luo2017placement}, etc. However, these solutions do not consider the convexification of the constraints. Consequently, they are unable to guarantee a global optimum. 








The contributions of this paper include three points. Firstly, this paper quantifies the protection device upgrade cost with step functions and integrates the protection cost into the objective function of the EV charging station placement. The effect of protection and voltage regulation upgrade on the charging station placement is revealed. Secondly, the convexification preservation is realized in this optimization problem, at the same time, the global optimum is achieved and guaranteed. Thirdly, this paper suggests a comprehensive sensitivity analysis before and after the problem convexification. The sensitivity validation further indicates the applicability of the proposed method in different cities and countries.



The established optimization problem originates from the practical concerns within electrical and transportation networks. Its sensitivity is firstly analyzed, then the constraint convexification is conducted. Meanwhile, the sensitivity analysis is re-evaluated after the problem convexification to see if it still holds. In the end, the proposed objective function along with its constraints will provide the results for the EV charging station planning, which satisfies the economic requirements that both networks request. The outline of the paper comes as follows: Section \ref{sec:problemformulation} elaborates on the mathematical formulation of the problem under study. Based on the proposed formulas, Section \ref{sec:probconvexEV} suggests a way of convexifying the proposed realistic constraints in the objective function. Section \ref{sec:evnumericalresults} demonstrates the numerical results as well as sensitivity analysis after convexification in small and large scale systems respectively. Furthermore, the discussions on the geographical effect, the importance of protection cost, and method expandability are presented in Section \ref{sec:EVplacementDiscussion}. The conclusions are in Section \ref{sec:conclusionsEVplacement}.


\begin{figure*}[!htb]
	\centering
	\includegraphics[width=6.7in]{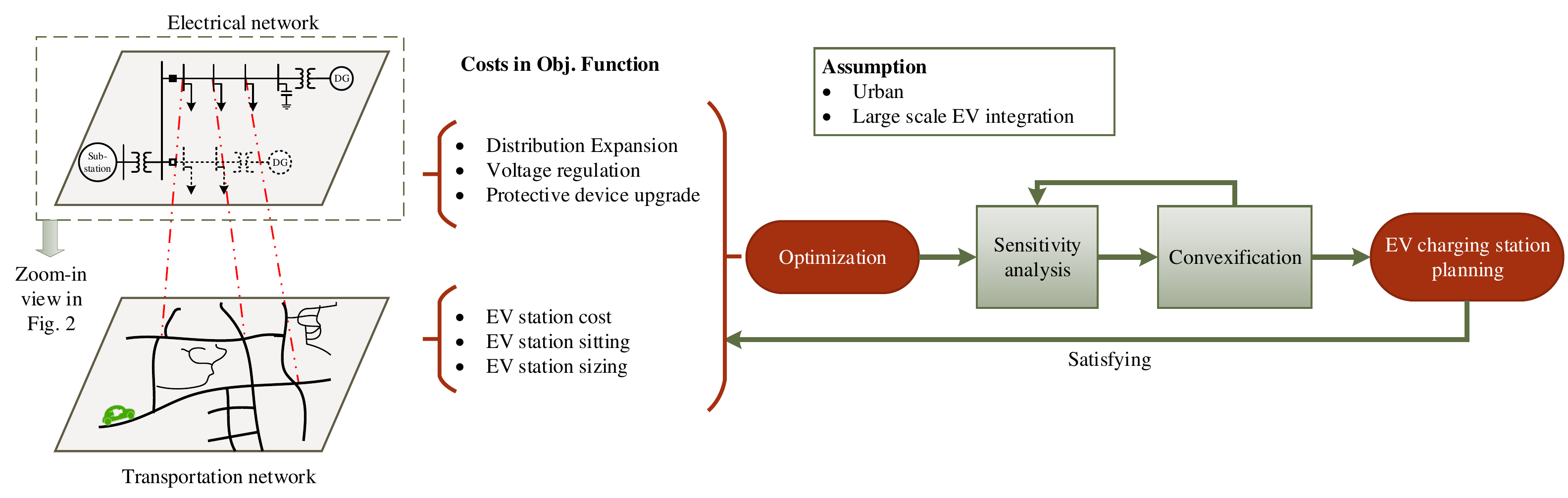}
	\centering
	\caption{Flowchart of the proposed EV charging station placement method.}
	\label{fig:fcEV}
\end{figure*}

\section{Problem Formulation}
\label{sec:problemformulation}


Fig. \ref{fig:fcEV} shows the flowchart of the proposed EV charging station placement method. It considers the integrated electrical and transportation networks as well as their associated infrastructure costs. Therefore, the costs related to distribution expansion, voltage regulation, protective device upgrade, and EV station construction are incorporated in the objective function. This study is assumed to be conducted for urban cities and large scale of EV integration in the future. Since a great amount of stations has to be installed in this circumstance, the EV charging station integration point could be at any bus along the distribution feeder as long as the operation constraints permit. In this section, the objective function and constraints are first formulated and then explained in details. Afterwards, the sensitivity analysis is provided from a mathematical angle. 






\subsection{Objective Function and Constraints}
\label{sec:objfuncconst}
The objective function minimizes the total cost among the costs of charging stations, distribution network expansion, voltage regulation and protection device upgrade. It is formulated as a mixed-integer nonlinear optimization problem: 



\begin{equation}
\underset{x_i,y_i}{\text{minimize}}\quad C_{sta}+C_{dis}+C_{vr}+C_{prot}
\label{eq:objfun}
\end{equation}

\begin{subequations}
\text{subject to}\quad
\begin{align}
& x_i \in\ \{0,1\}, \; i\in\Phi,\label{eq:subeq1}\\
& y_i \in \mathbb{Z}^+, \; i\in\Phi,\label{eq:subeq2}\\
& \sum_{i\in\Phi} g(y_i)\geq S, \; i\in\Phi,\label{eq:subeq3}\\
& f(V_i,\delta_i,P_i,Q_i) = 0, \; i\in\Phi,\label{eq:subeq4}\\
& |I_j|<I_{max,j}, j\in\Psi,\label{eq:subeq5}\\
& V_{min,i}<|V_i|<V_{max,i}, \; i\in\Phi,\label{eq:subeq6}\\
& C_{sta}+C_{dis}+C_{vr}+C_{prot} \leq C_{budget}, \label{eq:subeq7}
\end{align}
\label{eq:constraint_misc_ACpf}
\end{subequations}

\noindent where 

%
%
%
%
%
%

\begin{align} 
C_{sta} &=  \sum_{i\in\Phi} (c_{1,i}x_i+c_{2,i}y_i), \label{eq:stationcost} \\
C_{dis} &=  \sum_{i\in\Phi} (c_{3,i}l_i (P^{line}_{0,i}+\Delta P^{line}_i))+c_{4,i}h(\Delta P_i^{sub}), \label{eq:disexpansioncost}\\
C_{vr} &=  c_{5} \sum_{i\in\Phi}\Delta V_i^2, \label{eq:voltagereg}\\
C_{prot} &=  C_{acq}+C_{inst}+C_{uninst}+C_{main}. \label{eq:protupgradecost}
\end{align}

\noindent The definition of the notations can be found in Table \ref{tab:notations}. In the remaining part of Section \ref{sec:problemformulation}, the objective function and constraints are explained first, then the problem sensitivity is analyzed from the mathematical perspective. 

\begin{table}[!hbt]
	\renewcommand{\arraystretch}{1.3}

	\caption{Notation and parameter definition.}
	\label{tab:notations}
	\centering
	\begin{tabular}{>{\centering\arraybackslash}m{1.15cm} >{\centering\arraybackslash}m{6cm}}
		\hline
		\hline
		Notation & Definition\\
		\hline
		$\Phi$ & Set of nodes in the distribution networks.\\
		$\Psi$ & Set of branches in the distribution networks.\\
		$n$ & Total number of nodes in the distribution networks.\\
		$m$ & Total number of branches in the distribution networks.\\
		$x_i$ & Binary variable denoting charging station location at node $i$.\\
		$y_i$ & Number of new charging spots to be installed in the station of node $i$.\\
		$l_i$ & Length of the distribution line expansion required for a new charging station at node $i$.\\
		$P^{line}_{0,i}$ & Original line capacity at node $i$, in kVA.\\		
		$\Delta P^{line}_i$ & Expanded line capacity at node $i$, in kVA.\\		
		$\Delta P^{sub}_i$ & Expanded substation capacity at node $i$, in kVA.\\	
		$\Delta V_i^2$ & Voltage deviation from nominal voltage $V_{nom}$ at node $i$, in p.u.\\
		$I_{max,j}$ & Maximum current at branch $j$, in Amp.\\
		$V_{max,i}$ & Maximum voltage at node $i$, in Volt.\\
		
		$c_{1,i}$ & Fixed cost to build a new station at node $i$.\\
		$c_{2,i}$ & Fixed cost to add an extra spot in the existing charging station at node $i$.\\
		$c_{3,i}$ & Line cost at node $i$, in $\$/($kVA $\cdot$ km).\\
		$c_{4,i}$ & Substation expansion cost at node $i$, in $\$/$kW.\\
		$c_{5}$ & Voltage regulation cost coefficient per p.u. voltage square, in $\$/$ $p.u.^2$.\\
        $S$ & Number of charges in a certain area, aggregated by zip code.\\
		$p_{ev}$ & The power of the integrated EV, in $W$.\\
		\hline
		\hline
	\end{tabular}
\end{table}

\subsection{Explanations of the Objective Function}
\label{sec:expofobjfunc}

\begin{figure}[!htb]
	\centering
	\includegraphics[width=3.5in]{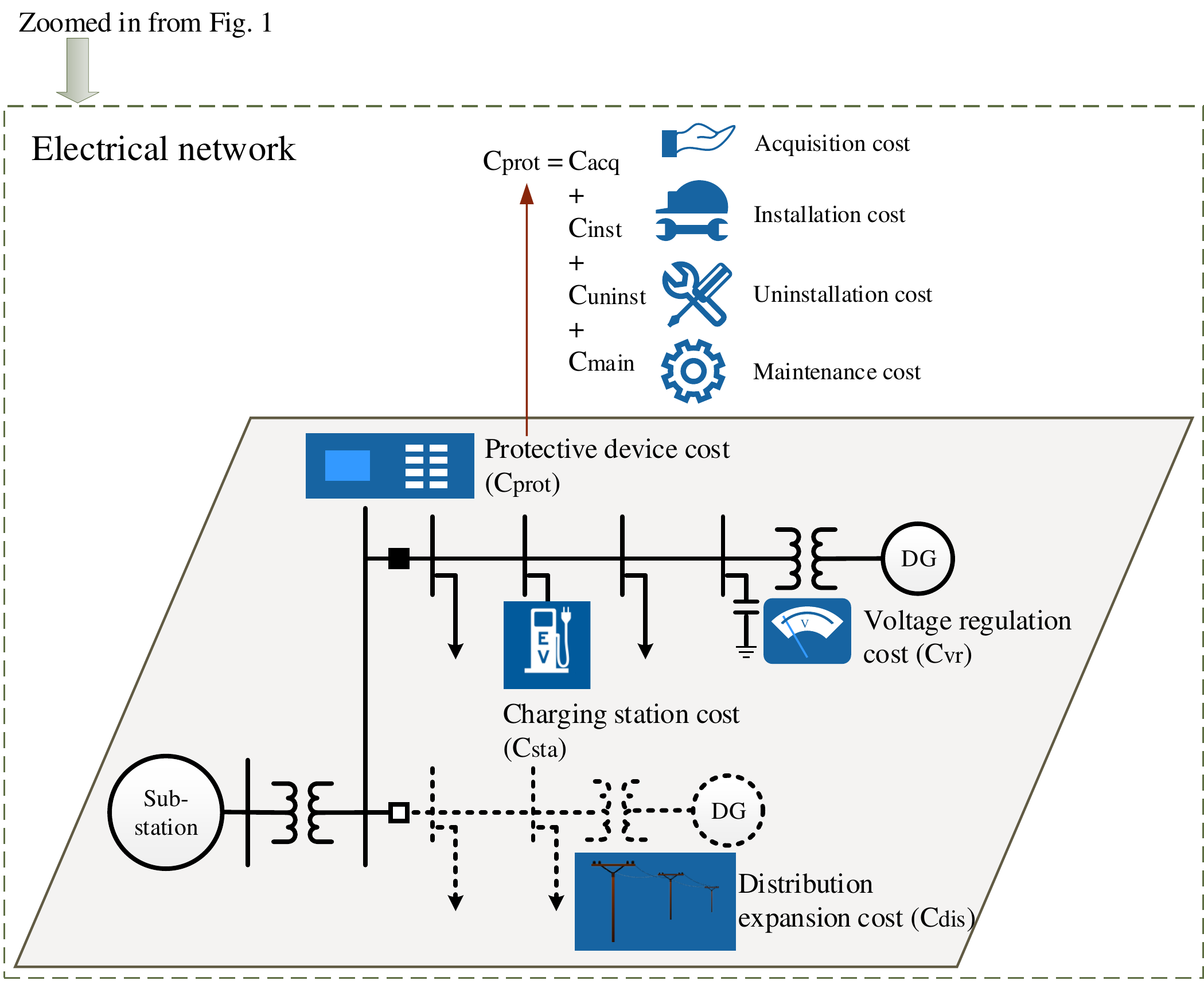}
	\centering
	\caption{Cost decomposition -- Zoom-in of the electrical network.}
	\label{fig:protectioncosticon}
\end{figure}

The objective function aims to minimize the total cost associated with four aspects in (\ref{eq:objfun}). They are visualized in Fig. \ref{fig:protectioncosticon}, which is zoomed in from the electrical network in Fig. \ref{fig:fcEV}. There are four terms (also viewed as four constraints)  in this objective function. It is actually a metric not only in planning but also in operation for the trade-off. 


The first term is the fixed cost of building a new station and of adding an extra spot in the existing charging station. The second one is related to the distribution line cost and the substation expansion cost. The distribution line cost is approximately proportional to the product of the line length and the total line capacity. Moreover, the total line capacity is the sum of the original line capacity and the expanded line capacity, where the latter can be estimated to be proportional to the number of new charging spots to be installed:

\begin{equation}
P^{line}_{0,i}+\Delta P^{line}_i=P^{line}_{0,i}+p_{ev}y_i, \quad i\in\Phi.
\label{eq:plinei}
\end{equation}

Also in the second term, a 1 MW substation surplus capacity is assumed, which makes the $\Delta P_i^{sub}$ a piecewise function:

\begin{equation}
h(\Delta P_i^{sub})=
\begin{cases} 
      0, & \Delta P_i^{sub}<P_{sur}, \\
      p_{ev}\sum_i y_i, & \Delta P_i^{sub} \geq P_{sur}.
\end{cases}
\label{eq:substationsurpluspower}
\end{equation}

\noindent In addition, the expanded line capacity at node $i$ is defined as

\begin{equation}
\Delta P^{line}_i=p_{ev}y_i,
\label{eq:plinepevyi}
\end{equation}

\noindent where $p_{ev}$ is the power of integrated EVs. 

The third term represents the equivalent cost resulting from the impact of EV charging stations on the distribution network voltage profile, where $\Delta V_i=V_i-V_{i,ref}$. \cite{ref:dukpa2011fuzzy} proposes a stochastic capacitor planning formulation for distribution systems. However, voltage regulation techniques in distribution systems include voltage regulation transformers, static var compensators, static synchronous compensators, and shunt capacitor banks, etc \cite{ref:kang2017interconnection}. They can maintain voltage levels of load buses within an acceptable range. Since $Q=V^2/X_c=\omega C V^2$, the amount of reactive power compensation is proportional to the bus voltage. Therefore the square of voltage deviation from reference voltage at each bus is employed to evaluate the voltage regulation related cost.


The fourth term is associated with the protection device upgrade due to the installation of EV charging stations. The protection cost decomposition is shown at the top of Fig. \ref{fig:protectioncosticon}. It is assumed that the acquisition, installation, uninstallation, and maintenance costs are constant at each operating current range. With the protective device type $d$ and capacity $c$, the matrix A indicates the total number of devices to be installed \cite{ref:meneses2013improving}:

\begin{equation}
A(d,c)=\sum_{l=1}^{m}(x_{ldc}-x_{ldc}^{base}).
\end{equation}

Therefore, we have: 

%
%
%

\begin{align} 
C_{acq} &=  \sum_{d=1}^{4} \sum_{c=1}^{R_d} c^{acq}_{dc} \cdot z_{dc} \cdot A(d,c),\\
C_{inst} &=  \sum_{d=1}^{4} \sum_{c=1}^{R_d} \sum_{l=1}^{m} (c^{inst}_{dc} \cdot x_{ldc}  \cdot (x_{ldc}-x_{ldc}^{base})),\\
C_{uninst} &=  \sum_{d=1}^{4} \sum_{c=1}^{R_d} \sum_{l=1}^{m} (c^{uninst}_{dc} \cdot x_{ldc}^{base}  \cdot (x_{ldc}-x_{ldc}^{base})),\\
C_{main} &=  \sum_{d=1}^{4} \sum_{c=1}^{R_d} \sum_{l=1}^{m} (c^{main}_{dc} \cdot x_{ldc}),
\end{align}

\noindent where
\begin{equation}
x_{ldc}=
\begin{cases} 
      1, & I_0+\Delta I_{ld}=I_0+n_{dn\_ev}^l \cdot I_{ev}^0 \in c, \\
      0, & \text{otherwise}.
\end{cases}
\end{equation}

$x_{ldc}$ and $x_{ldc}^{base}$ are both binary variables. $x_{ldc}=1$ when line $l$ is determined to its upgrade protective device to type $d$ capacity $c$. $x_{ldc}^{base}$ when line $l$ initially has the protective device of type $d$ capacity $c$. $R_d$ is the number of ranges of the continuous current of protective device type $d$. $z_{dc}$ is a binary variable and is zero when the element $(d,c)$ of the matrix $A$ is negative. $c^{acq}_{dc}$, $c^{inst}_{dc}$, $c^{uninst}_{dc}$, and $c^{main}_{dc}$ are the acquisition, installing, uninstalling, and maintenance costs respectively, for the protective device of type $d$ capacity $c$. $n_{dn\_ev}^l$ is the number of EVs at downstream of line $l$. We assume there is no relocation during device upgrade since the original placement of the recloser was very likely determined by the reach of the feeder relays. Replacement of the recloser with directionality function is preferred in practice \cite{ref:doyle2002reviewing}.


\subsection{Explanations of the Constraints}
\begin{enumerate}
\item Optimization variables in (\ref{eq:subeq1}) and (\ref{eq:subeq2}): $x_i$ is a binary variable that indicates the availability of the charging station at bus node $i$, and $y_i$ is an integer variable that shows the number of charging spots at bus node $i$.

\item Charging serviceability constraint in (\ref{eq:subeq3}): the charging station serviceability needs to satisfy a predetermined number of charges, S, in a certain area. The charging serviceability is the summation of the serviceability function $g(y_i)$ over all charging locations. To simplify the problem, we use
\begin{equation}
g(y_i)=Dy_i,
\label{eq:chargeserviceability}
\end{equation}

\noindent where $D$ is the charging demand that each spot satisfies. We can also use queue theory \cite{ref:kimura1994approximations} to model the charging serviceability $g(y_i)$. 

\item Power flow constraints in (\ref{eq:subeq4}): the integration of EVs and locally distributed generations should respect the constraints of electric network. The function $f$ denotes the power flow equations. 

\item Line current constraint in (\ref{eq:subeq5}): the current flowing in each line should not exceed the maximum rated current of the line. 

\item Voltage limits in (\ref{eq:subeq6}): for the operation safety, the voltage range of $0.95$ $\sim$ $1.05$ is recommended in this study.

\item Budget limits in (\ref{eq:subeq7}): in some scenarios, the budget needs to be considered in the overall problem. 

\end{enumerate}

\subsection{Sensitivity Analysis of the Problem}
\label{sec:dominatingterms}

In order to analyze the sensitivity of each constraint in the objective function (\ref{eq:objfun}), the comparison is made in (\ref{eq:stationcost})-(\ref{eq:protupgradecost}). Specifically, $C_{sta}$ in (\ref{eq:stationcost}) can be rearranged into $c_{1,i}\sum_{i\in\Phi} x_i+c_{2,i}\sum_{i\in\Phi} y_i$; while $C_{dis}$ in (\ref{eq:disexpansioncost}) can be decomposed into the following format using 
(\ref{eq:plinei})-(\ref{eq:plinepevyi}) when $\Delta P_i^{sub} \geq P_{sur}$:


\begin{equation} \label{eq:cdisreaggranged}
C_{dis} = \sum_{i\in\Phi}c_{3,i} l_i P^{line}_{0,i}+(c_{3,i}l_i p_{ev} +c_{4,i} p_{ev})\sum_{i\in\Phi} y_i. 
\end{equation}

Now each term in the objective function seems more comparable:

\begin{itemize}
\item Constraint (\ref{eq:stationcost}): As for $C_{sta}$, its $\sum_{i\in\Phi} x_i$ part is much smaller than its $\sum_{i\in\Phi} y_i$ part when $c_{1,i}$ is equal to $c_{2,i}$, because the number of charging stations is always less than the total number of EV charging spots. 
\item Constraint (\ref{eq:disexpansioncost}): Between $C_{sta}$ and $C_{dis}$, the values associated with $\sum_{i\in\Phi} y_i$ depends on their coefficient $c_{2,i}$ and $(c_{3,i}l_i p_{ev} +c_{4,i} p_{ev})$. Furthermore, $C_{dis}$ has a relatively constant part $\sum_{i\in\Phi}c_{3,i} l_i P^{line}_{0,i}$ whereas $C_{sta}$ has the sub-term $c_{1,i}\sum_{i\in\Phi} x_i$ that relies on the optimized number of stations. 
\item Constraint (\ref{eq:voltagereg}):  $C_{vr}$ could grow relatively faster than other terms when $c_5$ is remarkable due to the square term in the summation. Meanwhile, the $\Delta V_i^2$ part in (\ref{eq:voltagereg}) indicates a strong relevance  with placements that boost up bus voltages. 
\item Constraint (\ref{eq:protupgradecost}): The sub-terms in $C_{prot}$ does not increase as fast as the ones with $\sum_{i\in\Phi} y_i$, and they mainly depend on the current-cost relationship as assumed in Section \ref{sec:expofobjfunc}. This means that the EV charging station placements resulting in high current absorption would increase the cost in this term. 
\end{itemize}

\section{Problem Convexification}
\label{sec:probconvexEV}

Following the sensitivity analysis of the previous section, this section discusses the way of convexifying the nonlinear terms in the objective function. Furthermore, the approximation error during convexification is discussed in the second part. Although the convex preservation contributes to some errors during optimization, what is gained is the guarantee of a global minimum solution in both small and large electric systems. 

\subsection{Convexify the Problem}

The linearization of the AC power flow in constraint (\ref{eq:constraint_misc_ACpf}) is depicted in Appendix \ref{sec:acpf_linear}. This paper does not focus on AC power flow linearization. In addition, other constraints in (\ref{eq:constraint_misc_ACpf}) are linear. Therefore, greater emphasis is to be placed on the constraints from the objective function.

\subsubsection{Constraint (\ref{eq:stationcost})} It is a linear combination of the number of stations and the number of spots. Therefore it is convex. 
\subsubsection{Constraint (\ref{eq:disexpansioncost})} The first part of this constraint is linear, whereas the second part of this constraint is not linear as indicated in (\ref{eq:substationsurpluspower}). However, the piece-wise linear function (\ref{eq:substationsurpluspower}) becomes linear when the assumed 1 MW substation surplus capacity is exceeded. It actually means that as long as there are more than $1 MW/0.044 MW \approx 23$ spots to be built downstream from the entire substation, this constraint is linear. This is not a strict requirement in a system larger than the toy example. 
\subsubsection{Constraint (\ref{eq:voltagereg})} In this constraint, the optimization variable $x_i$ is linearly related to the net active power injection $P_i$ at bus $i$ in power flow calculation: 

In this constraint, the variable $V_i$ is a nonlinear function of the optimization variable $x_i$, which is linearly related to the net active power injection $P_i$ at bus $i$ in power flow calculation:

\begin{equation}
P_{i,inj}=P_{i,gen}-P_{i,load}-x_i p_{ev}, \; i\in\Phi,
\label{eq:linearizePinjX}
\end{equation}

but the variable $V_i$ is a nonlinear function of the optimization variable $x_i$. Utilizing the AC power flow linearization technique in Appendix \ref{sec:acpf_linear}. We now can easily establish the linear relation between the optimization variable $x$ and the non-slack bus voltage $V_N$ by plugging (\ref{eq:linearizePinjX}) into (\ref{eq:linearizeACpf}). It is convex and a global optimum is guaranteed.

\subsubsection{Constraint (\ref{eq:protupgradecost})} Given the assumption of this constraint, the protection cost is actually a summation of four piece-wise step functions including the costs of acquisition, installation, uninstallation, and maintenance. Its typical curve is plotted in Fig. \ref{fig:protectioncost}. To linearize the step functions, these step functions in Fig. \ref{fig:protectioncost} are approximated by three linear lines (the dash-dot lines in blue) using the linear curve-fitting algorithm.

\subsection{Convexification Error Analysis}

The convexification error analysis is conducted in the same order as in the previous section. The convexification of the AC power flow in constraint (\ref{eq:constraint_misc_ACpf}) uses the same linearization technique as the one in constraint (\ref{eq:voltagereg}) from the objective function. The following convexification errors are elaborated. 

\begin{itemize}
\item Constraint (\ref{eq:stationcost}): No approximation error associated with this constraint since it is a linear constraint itself. 
\item Constraint (\ref{eq:disexpansioncost}): The largest approximation error occurs at the turning point where 23 spots are planned but substation expansion is not yet required. However, if the number of spots to be built is larger than $23$, there will be no error associated with this constraint. 


%
%

\item Constraint (\ref{eq:voltagereg}): The approximation error originates from the quadratic term that is neglected in the derivation of the voltage-power equation in Appendix \ref{sec:acpf_linear}. The error in complex power originates from the high order series of the following Taylor expansion. If we neglect high order terms and defining $V=1-\Delta V$, a linear form is obtained when $||\Delta V|| <1$:

\begin{equation}
\frac{1}{V}=\frac{1}{1-\Delta V}=\sum_{n=0}^{+ \infty}(\Delta V)^n \approx 1+\Delta V=2-V. 
\label{eq:combinedVderive}
\end{equation}


The error in percentage for the approximation is calculated by defining a function $\Psi(V)=100 \cdot ||(1/V)-(2-V)||$. $L_2$-norm is employed here.

\item Constraint (\ref{eq:protupgradecost}): To simplify this constraint, a best-fitting straight line in Fig. \ref{fig:protectioncost} for each protective device is obtained based on realistic costs (refer to Appendix \ref{sec:protectioncostdistributionsystem}) and R-squared values as shown in Table \ref{tab:trendlineprotdeviceacquisition}. The closer the R-squared value is to 1.0, the better the fit of the regression line. We can see that all of the four types of devices' $R$ values are above $0.75$, which fairly represents the realistic device costs.  

\begin{figure}[!htb]
	\centering
	\includegraphics[width=3in]{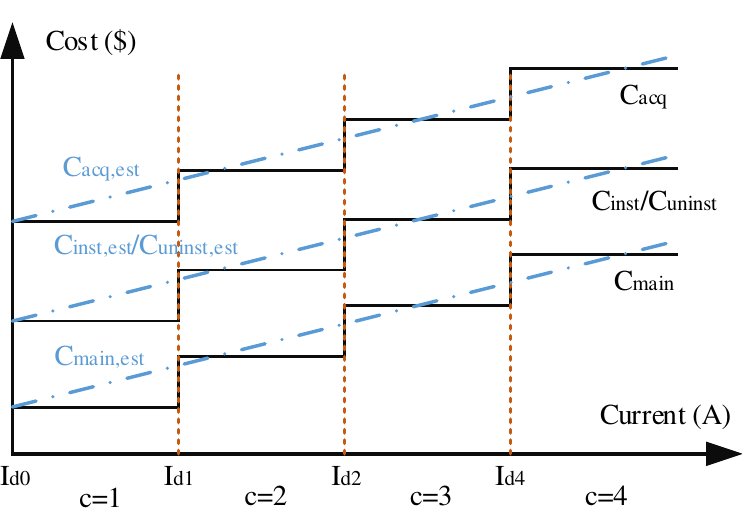}
	\caption{Cost functions of protective devices and their estimated and linearized functions. Assume there are four levels of capacity for the device type d.}
	\label{fig:protectioncost}
\end{figure}

\begin{table}[!hbt]
	\renewcommand{\arraystretch}{1.3}

	\caption{Best trend-line on estimating the protective device costs using the data in Table \ref{tab:protacq}.}
	\label{tab:trendlineprotdeviceacquisition}
	\centering
	\begin{tabular}{>{\centering\arraybackslash}m{2.5cm}>{\centering\arraybackslash}m{3cm}>{\centering\arraybackslash}m{1cm}}
		\hline
		\hline
		Device type & Cost function ($c$) w.r.t. current ($I$) & $R^2$\\
		\hline
		Fuse& $c = 3.771I + 548.26$ & $0.775$\\
		Recloser& $c = 16.381I + 18219$ & $0.797$\\
		Overcurrent relay& $c = 2.040I + 4515.9$ & $0.756$\\
		DORSR& $c = 2.040I + 6515.9$ & $0.756$\\
		\hline
		\hline
	\end{tabular}
\end{table}

\end{itemize}

%

\subsection{Sensitivity Analysis}
The errors according to the above analysis are small and have little influence on the sensitivity of the problem. For example, the constraint (\ref{eq:stationcost}) is linear by itself. The constraint (\ref{eq:disexpansioncost}) is also linear when below or above a certain number of charging spots. The constraint (\ref{eq:voltagereg}) has a maximum error of $0.26\%$ in the $\Psi(V)$ equation if the deployed voltage regulator regulates the bus voltage between $0.95$ and $1.05$. The constraint (\ref{eq:protupgradecost}) does not present a large cost change among each two adjacent current ranges based on the realistic data in Appendix \ref{sec:protectioncostdistributionsystem}. Therefore the protective cost can be assumed as a constant when the continuous current setting is within a range of operating currents \cite{ref:meneses2013improving}. In the next section, the numerical results will validate the problem sensitivity in various aspects. 

\section{Numerical Results}
\label{sec:evnumericalresults}

After discussing the way of convexifying the constraints in the objective function, this section demonstrates the effect of each constraint from the objective function on the overall problem using realistic data. After introducing the cost parameters and the systems under study, we investigate the sensitivity of the formulated problem in both small and large systems. 

\subsection{Cost Parameters and Systems Under Study}
\label{sec:costsandassumptions}

The utilized costs and assumptions in this section include:

\begin{itemize}
  \item The fixed costs for each PEV charging station is assumed to be $c_{1,i}=163,00$\SI{0}{(\$)} \cite{ref:zhang2017pev}. 
  \item The land use costs are \SI{407}{\$/m^2} and adding one extra charging spot requires \SI{20}{m^2} land. The per-unit purchase cost for one charging spot is $23,50$\SI{0}{\$} \cite{ref:agenbroad2014pulling}. Thus we have $c_{2,i}=407 \times20 + 23,500=31,64$\SI{0}{(\$)}.
  \item The distribution line cost is assumed to be $c_{3,i}=$\SI{120}{(\$/(kVA \cdot km))} \cite{ref:yao2016scenario}.
  \item The substation expansion cost is assumed to be $c_{4,i}=$\SI{788}{(\$/kVA}) \cite{ref:yao2014multiobjective}.
  \item The charging demand, $D$, that each spot satisfies, is assumed to be (\SI{24}{h}$\times$\SI{60}{min/h})$/$(\SI{42}{min}$\times 0.5$) = \SI{68}{(vehicles/day)}.\footnote{The average charging time of an EV with empty battery is estimated as (\SI{200}{km}$\times$\SI{0.14}{kWh/km})$/$(\SI{44}{kW}$\times 0.92$) = \SI{42}{min}.\cite{ref:zhang2016integrated}}
  \item We assume the distribution feeder has \SI{1}{MVA} surplus substation capacity which can be utilized by charging station.
  \item The rated charging power for each charging spot is \SI{44}{kW} \cite{ref:zhang2017pev}. Assume $c_{5}=$50,00$ $\SI{0}{(\$)}. Per car, the charging current is assumed to be \SI{44}{kW}$/ \sqrt{3}/$\SI{12.5}{kV}$=$\SI{2}{A}.
  \item  As for the protective devices, an overcurrent relay is assumed to be installed next to B2 and a fuse next to B3. Meanwhile, B2 and B3 are also the only buses where an EV charging station can be built.
  \item Assuming there are about $85$ EVs per hour require charging services in the area under study and there is no limit for each charging spot. Therefore, we will have the total charging station spots of $85 \times 24/68=30$. 
\end{itemize}

\begin{figure}[!htb]
	\centering
	\includegraphics[width=3.0in]{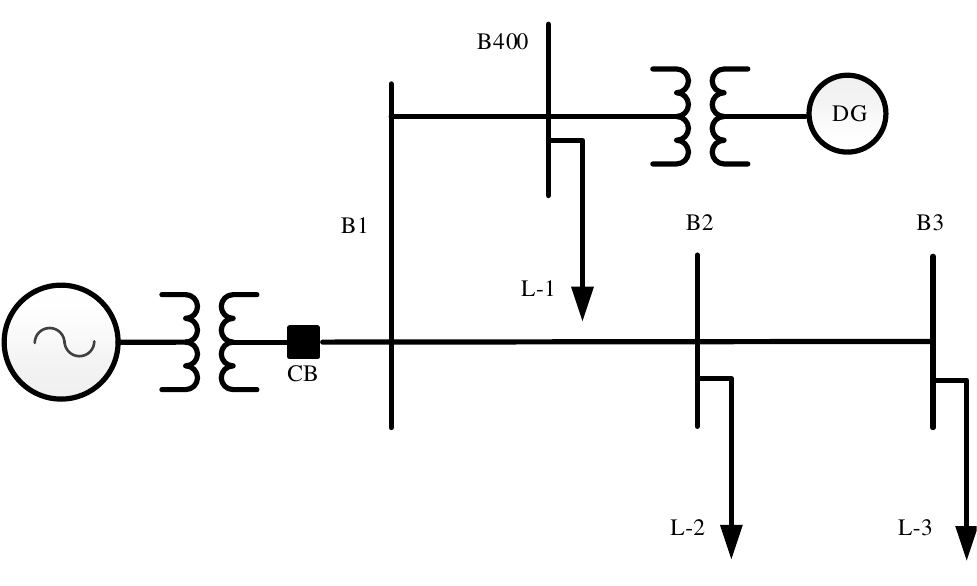}
	\caption{IEEE 4-bus distribution system.}
	\label{fig:4bussystem}
\end{figure}

The toy example is based on a modified IEEE 4-bus system as shown in Fig. \ref{fig:4bussystem}. The IEEE 123-bus distribution system is used for the large system study. 

\subsection{Numerical Results of a Toy Example}
\label{sec:toyexampleev}

This subsection demonstrates the results in an IEEE 4-bus small toy example then draws some interesting observations from this toy example. The results on large systems are presented in the next subsection. 

The placement results are presented by adding one constraint item after another to clearly observe the sensitivity of each constraint. Since there are many possible types of permutation of adding the four constraints, we have selected the constraint incremental procedure that best illustrates the nature of each constraint as shown in Table \ref{tab:toyexamplecost}. The following subsections illustrate the consequences in four representative scenarios:

\begin{table}[!hbt]
	\renewcommand{\arraystretch}{1.3}

	\caption{EV charging station placement results of the toy example by adding the constraints incrementally.}
	\label{tab:toyexamplecost}
	\centering
	\begin{tabular}{>{\centering\arraybackslash}m{3.3cm}>{\centering\arraybackslash}m{2.5cm}>{\centering\arraybackslash}m{1.2cm}}
		\hline
		\hline
		Constraint & EV numbers at B2 \& B3 & $C_{total} (\$)$\\
		\hline
		$C_{sta}$& $(0,30)$ or $(30,0)$ & $949,200$\\
		$C_{sta}+C_{dis}$& $(0,30)$ or $(30,0)$ & $2,152,360$\\
		$C_{sta}+C_{dis}+C_{vr}$ & $(0,30)$ & $2,160,360$\\
		$C_{sta}+C_{dis}+C_{vr}+C_{prot}$& $(30,0)$ & $2,195,210$\\
		\hline
		\hline
	\end{tabular}
\end{table}


\subsubsection{The Constraint Added: $C_{sta}$}

When there is only one constraint of charging station cost, the objective function attempts to build less number of stations for reducing the total cost as shown in (\ref{eq:stationcost}). Since no spot limitation is assumed in this example, the optimal placements are a) no station built near B2 and 30 spots built near B3, or b) no station built near B3 and 30 spots built near B2. They are noted as $(0,30)$, $(30,0)$.

\subsubsection{The Constraint Added: $C_{sta}+C_{dis}$}
When the constraint of distribution system expansion is included, the EV placement is not changed. The reason for that is the constraint of $C_{dis}$ depends on $\sum y_i$ -- the total number of spots, which adds more cost but does not alter the placements within the buses for station installation. It is more economical to build fewer stations since the cost is saved by building one charging station as long as the capacity of the station is not violated. To be noticed that the spot capacity can affect the total cost in this case. 
      

%

\subsubsection{The Constraint Added: $C_{sta}+C_{dis}+C_{vr}$}

Now the voltage regulation constraint is also added to the objective function. It plays an influential role in favor of the placement that causes less voltage deviation. The resulting placement of $(0,30)$ indicates that we have the minimum voltage violation by placing all the EVs at bus 3 (end of the feeder). 
To be noticed that voltage regulation cost might overwhelm other costs. 
  
 
%
 
\subsubsection{The Constraint Added: $C_{sta}+C_{dis}+C_{vr}+C_{prot}$}
With the last constraint from a protection upgrade perspective, the optimal placement becomes $(30,0)$. With the existence of this constraint, the placement moves from the end of the feeder towards the substation due to the characteristics associated with protective device upgrade: (a) the overcurrent relay (OC) relay upgrade at branch 12 (1 and 2 are the from and to buses respectively) is inevitable; (b) if all of the EV spots are placed at the end of the feeder, more protective devices are required to be upgraded along the feeder, and in this case, case 2 costs less since protective devices at branch 23 do not require an upgrade. Additionally, the land cost coefficients can overwhelm the voltage regulation cost and the protection cost, if the land cost in the urban area is expensive.   

%

%

\subsection{Numerical Results on a Large System}
\label{sec:numericalresultEV}

The numerical results in the previous subsection exhibit the way each constraint affects the placements from the small-scale system perspective. This subsection reveals the sensitivity analysis after convexification preservation on large systems. The benefits of problem convexification are first discussed. To observe the placement results, the optimization variables are evaluated. Subsequently, the cost analysis is added to validate the sensitivity. The deployed benchmark system in this section is the IEEE 123-bus distribution system, and the costs and assumptions follow the ones in Section \ref{sec:costsandassumptions}.

\subsubsection{The Benefits of Problem Convexification}

Efforts are exerted on the convexification of the nonlinear constraints, the purpose of which is to guarantee a global optimum without jeopardizing the cost evaluation. Table \ref{tab:optimwithwithoutconvex} illustrates the comparison between the scenario that convexifies all the constraints and the one does not. 

%

\begin{table}[!hbt]
	\renewcommand{\arraystretch}{1.3}

	\caption{Optimization results with and without constraint convexification in the 123-bus System.}
	\label{tab:optimwithwithoutconvex}
	\centering
	\begin{tabular}{>{\centering\arraybackslash}m{3.5cm}>{\centering\arraybackslash}m{1.8cm}>{\centering\arraybackslash}m{1.8cm}}
		\hline
		\hline
		Constraints & Without convexification & With convexification\\
		\hline
		Percent of cases that failed to find a global minimum & $19.8\%$ &  $0.0\%$ \\
		Average total cost in cases with a global minimum (\$) & $7.92\times 10^7$ & $7.96\times 10^7$ \\
		Average total cost in cases with a local minimum (\$) & $8.47\times 10^7$ & Not available \\
		Computational time in cases with a global minimum (sec) & $112.4$ & $105.6$ \\
		Computational time in cases with a local minimum (sec) & $2,115.7$  & Not available \\
		\hline
		\hline
	\end{tabular}
\end{table}

Firstly, there are $23$ cases tested in this section under different EV flows and station capacity limits in order to obtain the percent of cases that failed to find a global minimum. As is seen from Table \ref{tab:optimwithwithoutconvex}, $19.8 \%$ of cases failed to find a global minimum due to the non-convexity constraints. Not surprisingly, all of the cases with convexified constraints successfully find the global minimum. Secondly, the fact of convexifying the constraints does not affect much of the total cost. As shown in the table, convexifying the constraints introduces only $(7.96\times 10^7-7.92\times 10^7)/7.92\times 10^7=0.51\%$ of error on the total cost, given the EV flow of 5185 EVs/h and 25-spot limit per station (same conditions are applied in the remaining row of Table \ref{tab:optimwithwithoutconvex}). Thirdly, the total optimization computational time with convexified constraints is comparable with the one without convexification. However, the computational time is significantly high in the cases that a local minimum is found. In summary, the convex preservation contributes to a limited amount of extra cost to the total cost and always provides a global minimum with small computational time. Therefore it is concluded that the idea of convexifying the constraint in this problem is more beneficial than disadvantageous in this optimization problem.

\subsubsection{Sensitivity Validation With Respect to Optimization Variables}
\begin{figure}[!htb]
	\centering
	\includegraphics[width=3.5in]{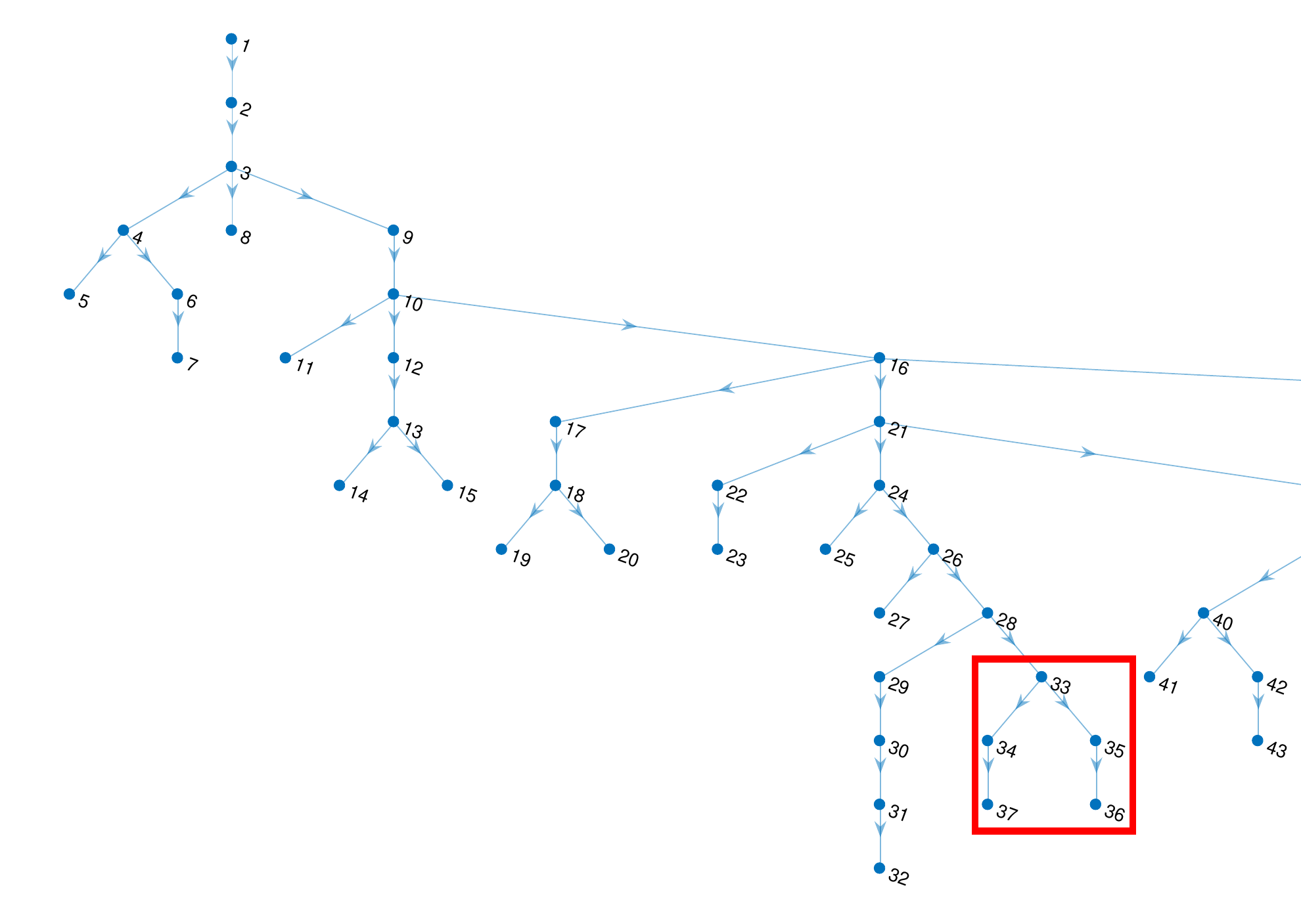}
	\centering
	\caption{The partial topology of the distribution system and the highlight of the area under study.} 
	\label{fig:bus123topopartial}
\end{figure}

Part of the system's topology is shown in Fig. \ref{fig:bus123topopartial}. Let us assume that the spot limitation of each station in this network is $25$. Here, we focus on five particular nodes at bus 33-37 to see how the constraints affect the EV charging station placement. Table \ref{tab:case123bus3337} shows the resulting placement as the constraint coefficient changes. Assuming the number of EV flows per hour requires only $30\%$ of the maximum station capacity in the whole system. When the voltage regulation constraint is not playing a role in the planning, due to the low density of the EV integration in this case study, only $4$ spots are required on bus $33$ to $37$. As the voltage regulation coefficient increases from $1e4$ to $5e5$, the total EV spot number over the small region of bus $33-37$ increases from $21$ to $50$, if we sum up the spot numbers of the second and third data rows in Table \ref{tab:case123bus3337}. This actually means when the voltage regulation cost is high, the preferred EV placement location moves to this region. As the voltage regulation coefficient goes higher, each bus in this region reaches its maximum capacity. Furthermore, the last data row in Table \ref{tab:case123bus3337} indicates the domination of the voltage regulation constraint does not rely on the existence of the protection constraint. 

\begin{table}[!hbt]
	\renewcommand{\arraystretch}{1.3}

	\caption{EV charging station placement at bus 33-37 in 123-bus system.}
	\label{tab:case123bus3337}
	\centering
	\begin{tabular}{>{\centering\arraybackslash}m{1.8cm} >{\centering\arraybackslash}m{3.2cm} >{\centering\arraybackslash}m{2cm}}
		\hline
		\hline
		Constraint & Constraint coefficient & Placement at bus 33-37\\
		\hline
		(\ref{eq:stationcost}), (\ref{eq:disexpansioncost}), (\ref{eq:protupgradecost}) & $c_5=0$ & 0 4 0 0 0\\ 
		(\ref{eq:stationcost}), (\ref{eq:disexpansioncost}), (\ref{eq:voltagereg}), (\ref{eq:protupgradecost})  & $c_5=1e4$ & 10 5 4 1 1\\
		(\ref{eq:stationcost}), (\ref{eq:disexpansioncost}), (\ref{eq:voltagereg}), (\ref{eq:protupgradecost})  & $c_5=5e5$ & 5 16 4 0 25\\
		(\ref{eq:stationcost}), (\ref{eq:disexpansioncost}), (\ref{eq:voltagereg}), (\ref{eq:protupgradecost})  & $c_5=1e6$ & 25 25 25 25 25\\
		(\ref{eq:stationcost}), (\ref{eq:disexpansioncost}), (\ref{eq:voltagereg}) & $c_5=1e6$ & 25 25 25 25 25\\
		\hline
		\hline
	\end{tabular}
\end{table}

By observing the overall placement results in the 123-bus system, the following conclusions are drawn:





\begin{itemize}
\item The constraint on voltage regulations pushes the EV charging station placement towards the end of the distribution feeder.
\item The cost derived from the constraint on the protective device is less when the EV charging stations are located near the feeder trunk.
\end{itemize}

In sum, the aforementioned observations along with the ones from the toy example are aligned with the sensitivity analysis of the objective function in Section \ref{sec:dominatingterms}. 


\subsubsection{Sensitivity Validation With Respect to Different Cost Components}

\begin{figure*}[htb!]
     \centering
     \subfloat[EV stations and spots. EV station capacity 25.]{\label{fig:placement25limit1}\includegraphics[width=3.5in]{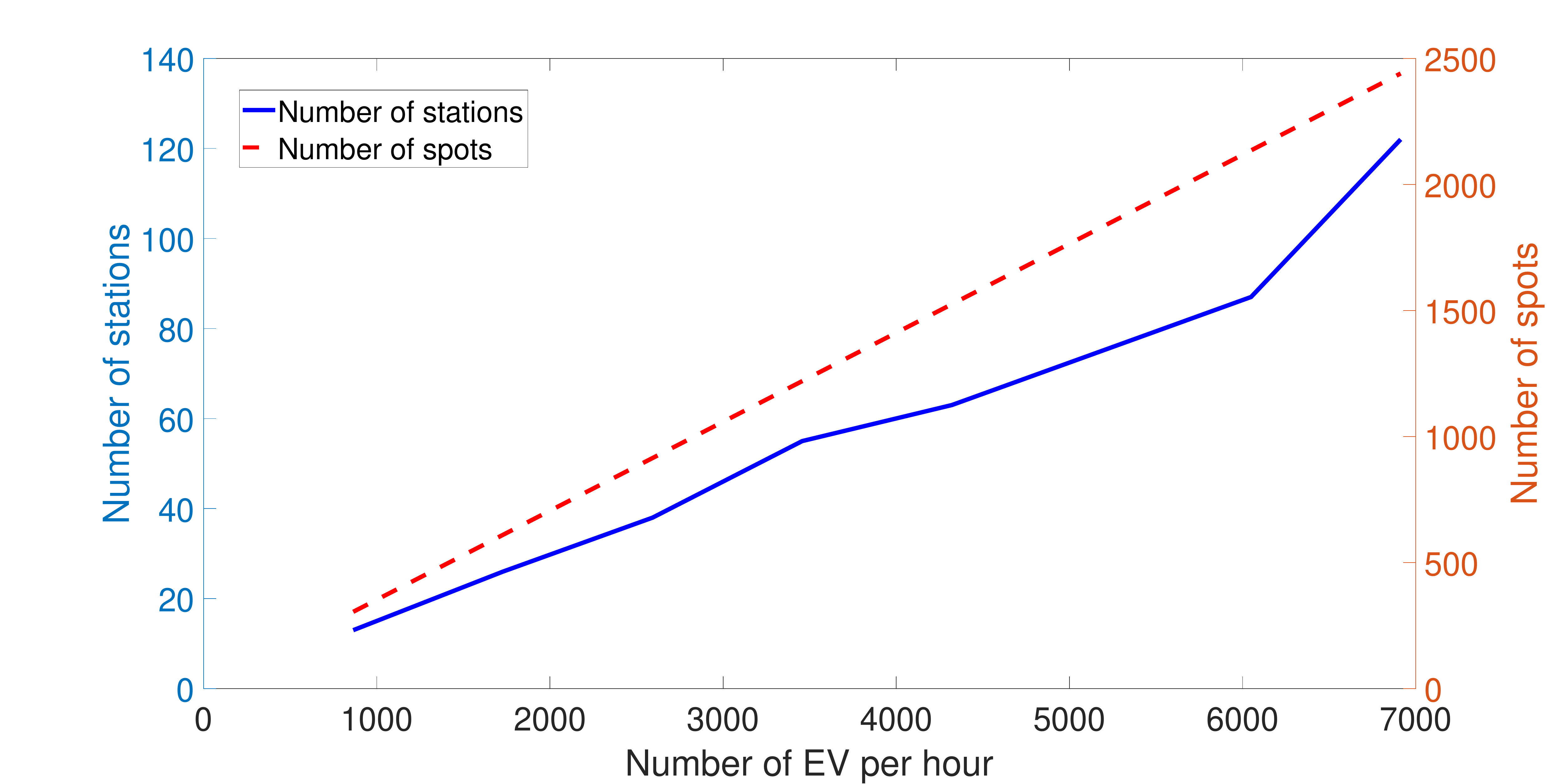}} 
     \subfloat[EV stations and spots. EV station capacity 10.]{\label{fig:placement10limit1}\includegraphics[width=3.5in]{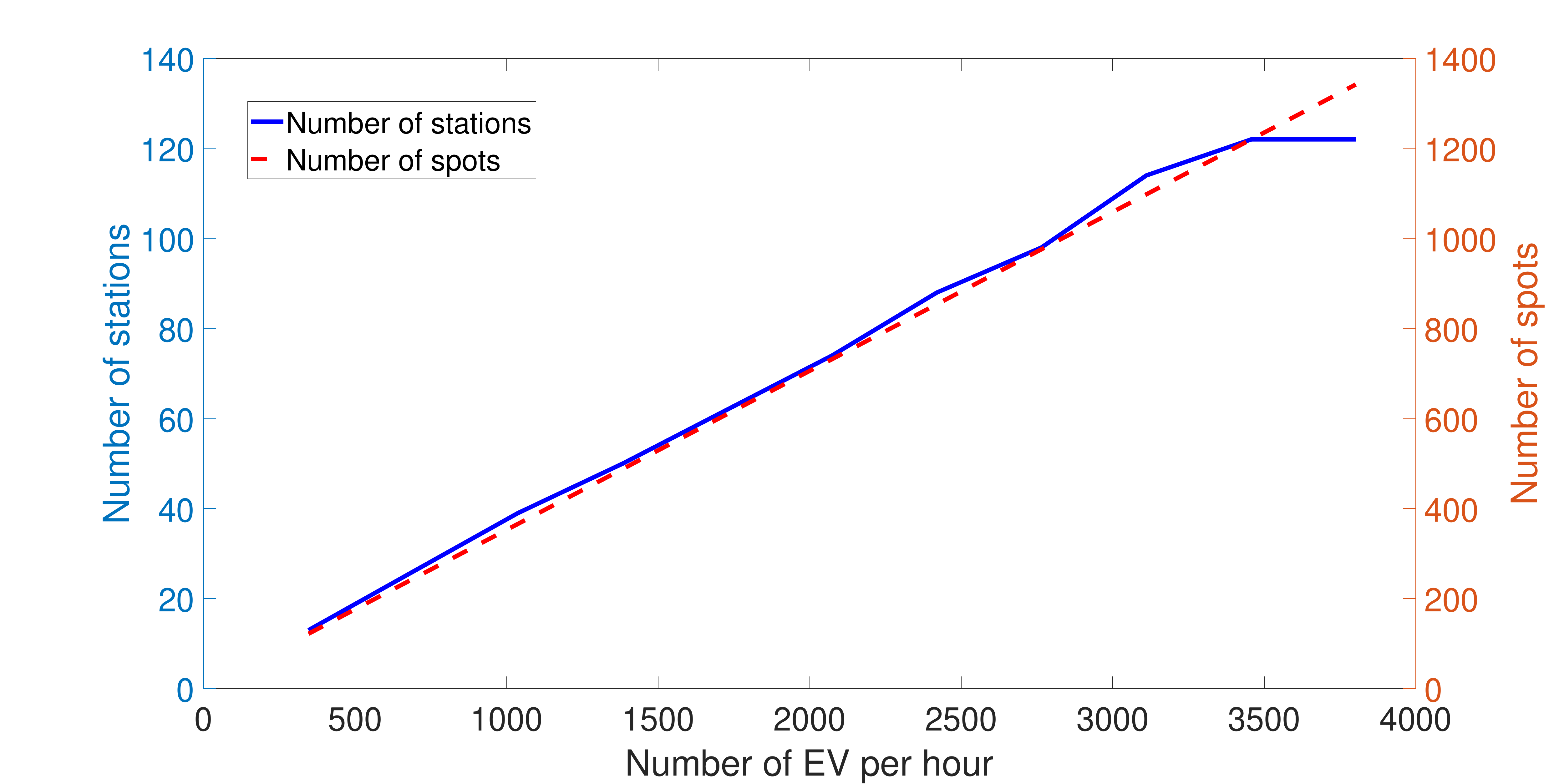}} \\
     \subfloat[EV costs. EV station capacity 25.]{\label{fig:placement25limit2}\includegraphics[width=3.5in]{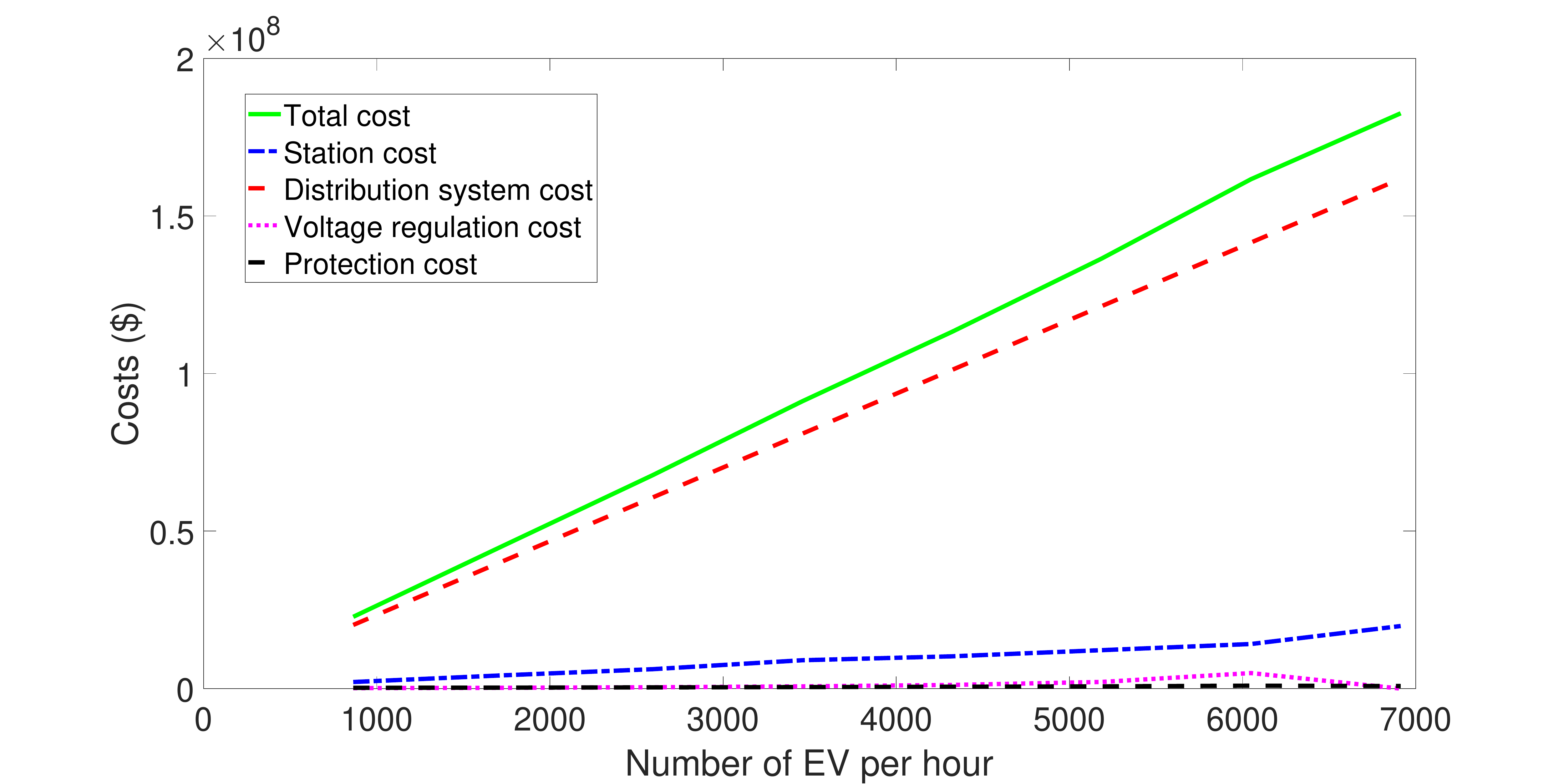}}
     \subfloat[EV costs. EV station capacity 10.]{\label{fig:placement10limit2}\includegraphics[width=3.5in]{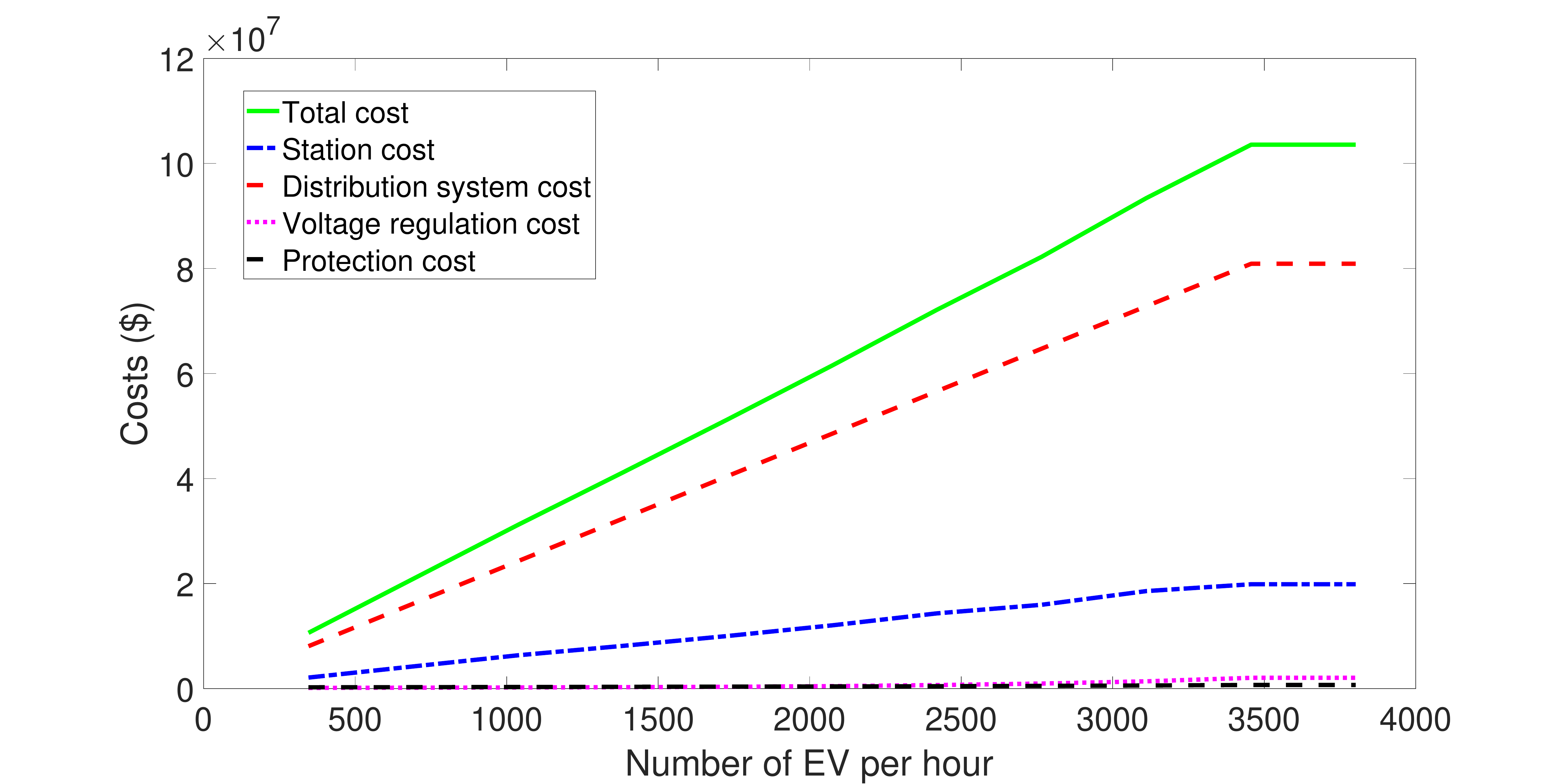}}
       \caption{EV charging station placement in the 123-bus system under different EV magnitude. $c_5=5e5$.} 
     \label{fig:placement25limit}
 \end{figure*}

In this subsection, we investigate three issues. First of all, how does the amount of EV flow at unit time affect the number of charging stations and total cost? As the number of charges per hour progresses, the number of spots in demand is proportional to the number of EVs per hour, as assumed and governed by (\ref{eq:chargeserviceability}). As for the number of stations, it reaches its maximum of $6,913$ in Fig. \ref{fig:placement25limit1}, which is bounded by the electric system capacity constraints (\ref{eq:subeq5}) and (\ref{eq:subeq6}). In Fig. \ref{fig:placement10limit1}, given the EV station capacity of $10$ spots per station, the number of stations saturates at $122$ -- the maximum number of stations that the current system can hold, when the number of EVs per hour reaches $3,500$. The cost diagrams under two EV station capacities are depicted in Fig. \ref{fig:placement25limit2} and Fig. \ref{fig:placement10limit2}. The cost of distribution system takes up a large portion of the total cost, whereas the costs of voltage regulation and protection device upgrade have low cost with the same parameters described in Section \ref{sec:toyexampleev}.

\begin{figure*}[htb!]
     \centering
     \subfloat[On the station number.]{\label{fig:c4change_stations}\includegraphics[width=3.5in]{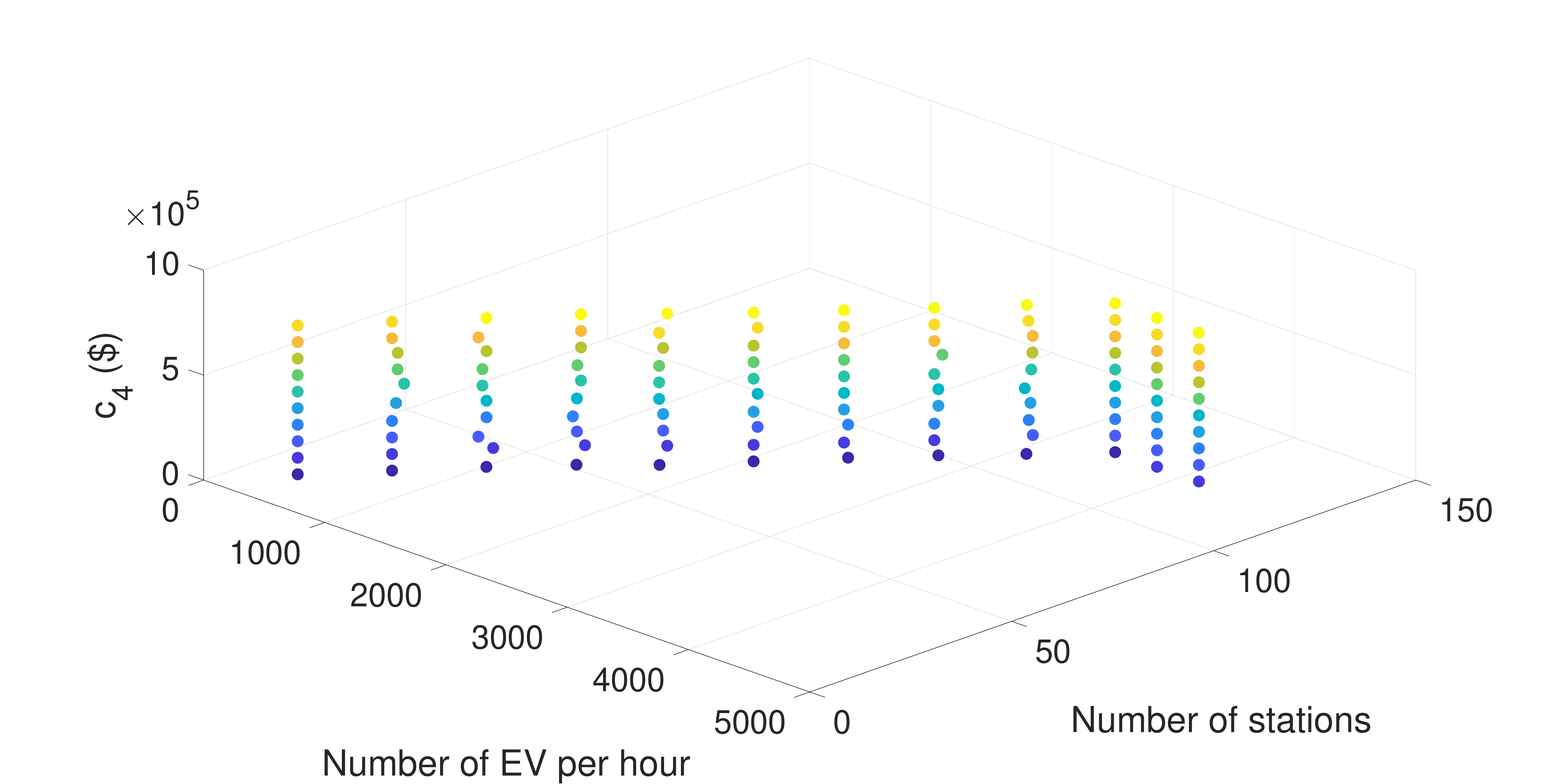}}
     \subfloat[On the total cost.]{\label{fig:c4change_totalcost}\includegraphics[width=3.5in]{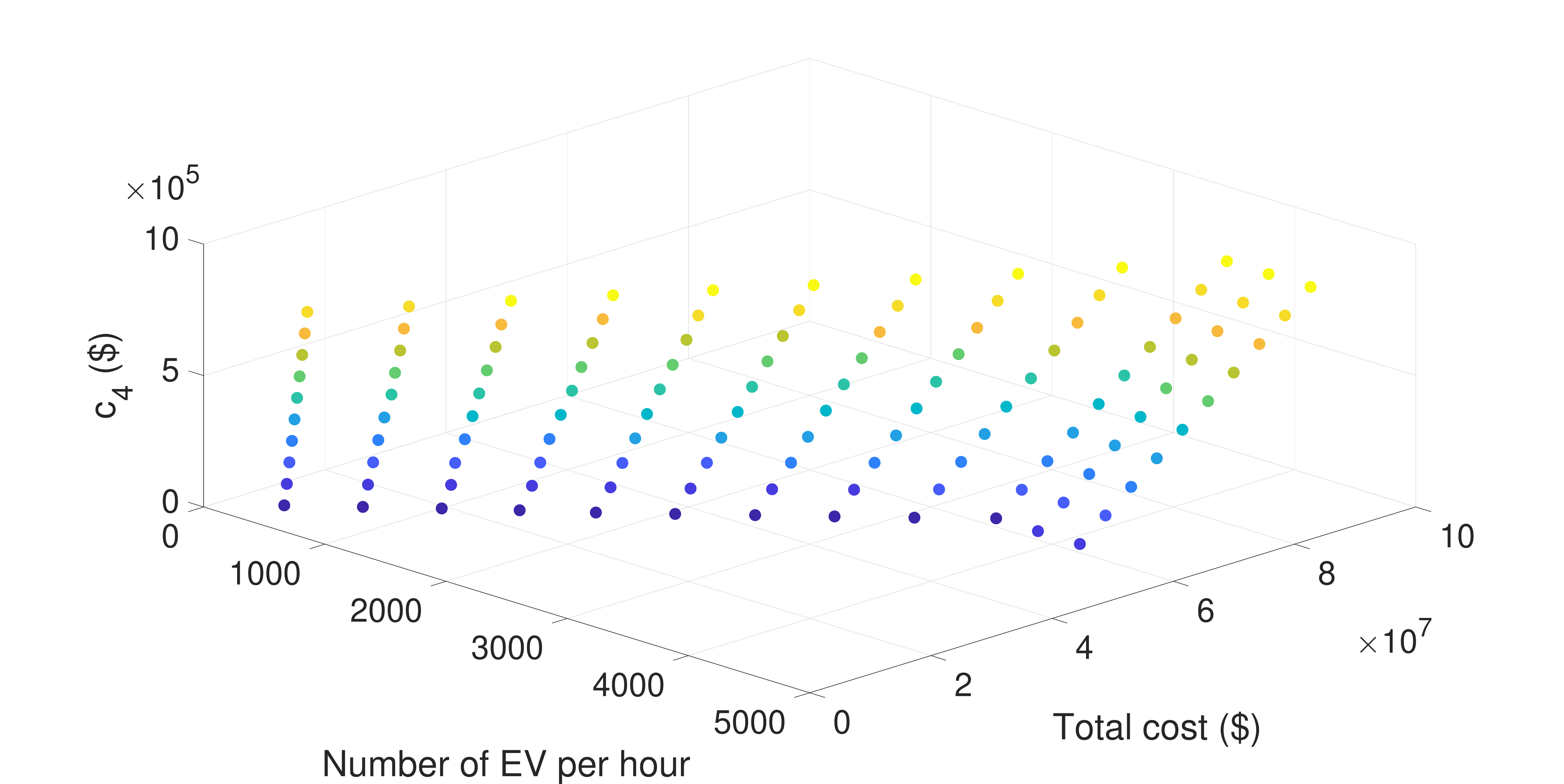}}
	\caption{The effect of the coefficient $c_4$ on the station number and the total cost. $c_5=5e5$. EV station capacity 10.} 
     \label{fig:c4change_sta_tot}
 \end{figure*}

Secondly, what is the effect of distribution expansion cost on the total cost? Due to the labor and land costs in different areas, costs resulting from (\ref{eq:stationcost}) and (\ref{eq:disexpansioncost}) varies immensely. Under this circumstance, the effect of the substation expansion coefficient $c_4$ on the EV charging station placements is investigated and plotted in Fig. \ref{fig:c4change_stations} and Fig. \ref{fig:c4change_totalcost}. From the bottom to top points, the same layers/color represents the same value of $c_4$. It can be observed that the number of stations does not rely on the varying of $c_4$. The increasing of $c_4$ does not change the planning of the stations but the total cost. It is easy to see that the larger number of EVs per hour there is, the more $c_4$ variation alters the total costs.



\begin{figure}[!htb]
	\centering
	\includegraphics[width=3.5in]{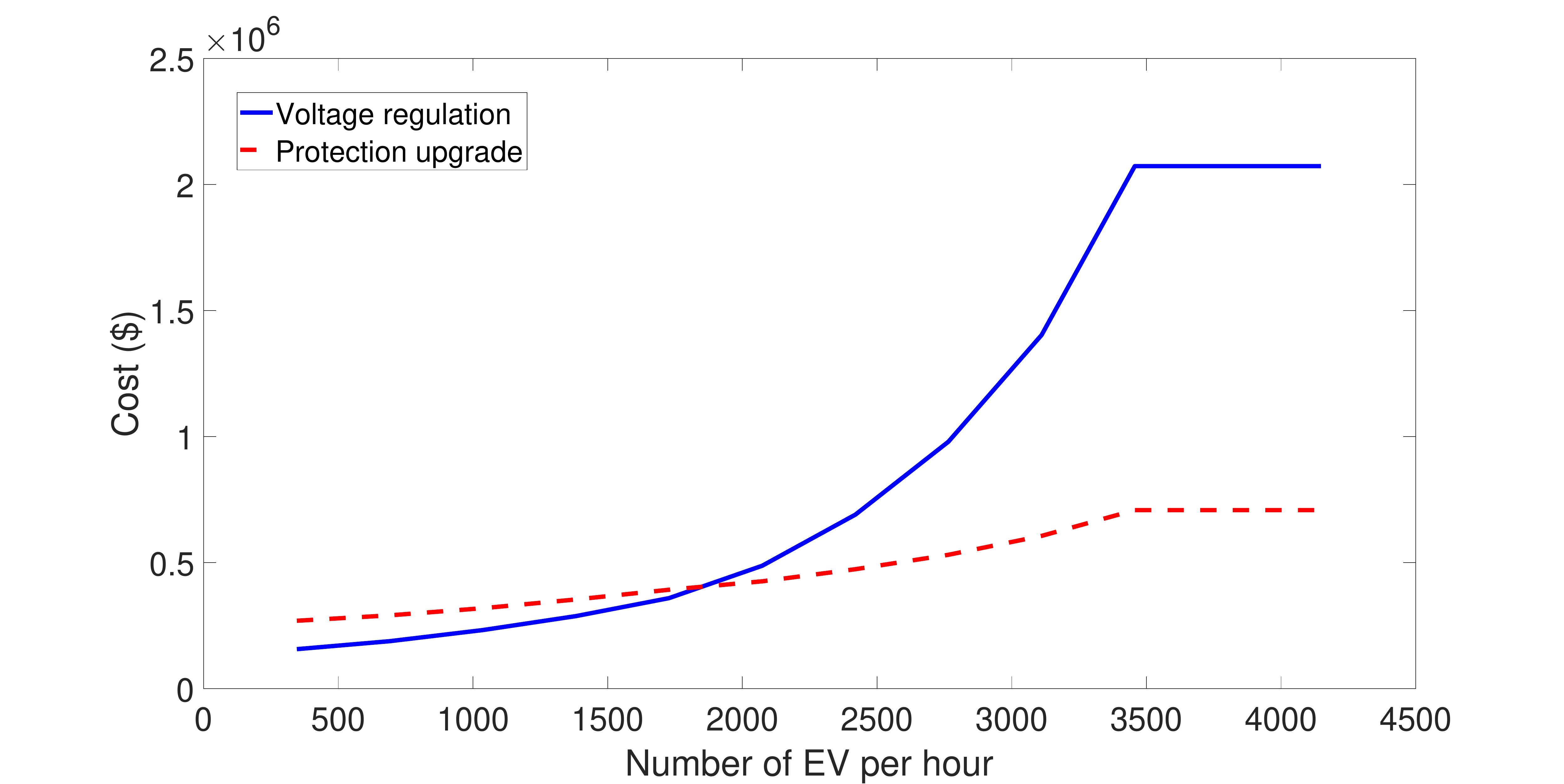}
	\centering
	\caption{The effect of the coefficient $c_4$ on the EV charging station placement. $c_5=5e5$. EV station capacity 10.} 
	\label{fig:c4change_vrprot}
\end{figure}

Thirdly, what are the relations between the amount of EV flow at unit time and the distribution grid operation costs on the voltage regulation and protection upgrade? When the EV charging station cost and distribution expansion cost are not dominantly high, the voltage regulation cost and projection cost affect the total cost. The sensitivity of the voltage regulation cost and projection cost in terms of the number of EVs per hour is presented in Fig. \ref{fig:c4change_vrprot}. The voltage regulation cost rises quadratically as predicted in Section \ref{sec:dominatingterms} and cease rising when the number of EVs per hour exceeds the system station capacity, which is $3,500$ EVs per hour. 

\section{Discussions}
\label{sec:EVplacementDiscussion}


The EV charging station placement is investigated in other distribution systems and scenarios. In this section, the geographical effect on the cost, the protection device cost, and the expandability of the proposed method are discussed. 

\subsection{Applicability in Different Cities and Countries}
\label{sec:cityncountry}

The formulation of the problem renders itself the flexibility of implementing different types of cost. Due to the variation of costs on land, labor, and equipment in different cities and countries, the coefficient $c_1$ to $c_5$ can vary significantly. According to the analysis in Section \ref{sec:dominatingterms}, the dominating terms are highly dependent on these coefficients. Although the aforementioned costs directly determine the coefficients of constraints (\ref{eq:stationcost})-(\ref{eq:protupgradecost}), the objective function remains effective in different locations because the objective function aims to minimize the total cost. Actually, in different cities or countries, the dominating constraints might be different. 

\begin{itemize}
\item Developed countries. Take the States, for example, the labor cost is comparatively high. According to \cite{ref:smith2015costs}, the labor cost takes up to $60\%$ of the EV supply equipment installation cost. That is why the EV charging station and distribution expansion costs are dominantly high in Table \ref{tab:toyexamplecost}. 
\item Developing countries. We investigate the EV charging station cost for the city of Beijing in China, the corresponding costs are listed, according to \cite{ref:li2011pricing}, as follows (assuming the US Dollar (USD) to Chinese Yuan (CNY) exchange rate is $6.35$): $c_{1,i}=50,64$\SI{0}{(\$)}, $c_{2,i}=7,12$\SI{2}{(\$)}, $c_{3,i}=$\SI{43}{(\$/(kVA \cdot km))}, $c_{4,i}=$\SI{102}{(\$/kVA}) It is seen that the EV charging station and distribution expansion costs are much lower compared to the US. The discrepancy in small cities in China is even larger. This phenomenon could eventually make the constraint of voltage regulation the dominant constraint during optimization. 
\item Regions where the price of the certain type of devices is significantly cheap or expensive. This case exists in some third world countries, where some devices are cheap because of its accessibility or expensive because of the import duty. In the toy example case study, when the coefficient of fuse cost in Table \ref{tab:trendlineprotdeviceacquisition} is dropped by two thirds due to the low cost in certain regions or countries, the optimal placement changes from $0,30$ to $30,0$ at bus $2$ and $3$ individually, meaning that the increasing of the weight of protection devices results in a strong constraint that pushes the EV chargers towards the substation and thus has less expense on protective device upgrade. 
\end{itemize}



\subsection{Results Without Considering the Protection Cost}

As the electric network grows larger, more protective devices are implemented. At the meanwhile, the cost of protective devices elevates even faster since the larger electric network can accommodate more EVs which result in higher steady-state and fault current levels. Consequently, in bigger systems, the cost of the protective devices is altering the EV charging station placement since we are minimizing the total operational cost. Table \ref{tab:largesystemEVplacement} illustrates the EV charging station placement results in two distribution systems under the EV flow of $1,500$ per hour with the parameters from Section \ref{sec:cityncountry}. It is presented that the protection cost may take up to $8\%$ of the total cost, which cannot be neglected in EV haring station placement planning. 

%

\begin{table}[!hbt]
	\renewcommand{\arraystretch}{1.3}

	\caption{EV charging station placement when the EV flow requires $30\%$ of available spots and the spots limit of each station is $10$.}
	\label{tab:largesystemEVplacement}
	\centering
	\begin{tabular}{>{\centering\arraybackslash}m{3cm}>{\centering\arraybackslash}m{1cm}>{\centering\arraybackslash}m{1cm}>{\centering\arraybackslash}m{1cm}}
		\hline
		\hline
		Sys. magnitude & 18-bus & 115-bus & 123-bus\\
		\hline
		\# of stations & $6$ & $39$ & $39$\\
		\# of spots & $51$ & $342$& $366$\\
		Total cost (\$) & $2.82e6$ & $1.86e7$& $5.30e6$\\
		Percent of prot. cost (\$) & $5.89\%$ & $8.60\%$& $6.03\%$\\
		\hline
		\hline
	\end{tabular}
\end{table}

\subsection{Expandability of the Planning Strategy}

The procedures of adding and convexifying additional constraints make the proposed optimization method suitable for implementing extra optimization constraints, meanwhile, it also guarantees a global optimum regardless of how large the system size is. The proposed method, therefore, becomes a feasible tool for EV charging station placement and planning.

\section{Conclusions}
\label{sec:conclusionsEVplacement}

This paper formulated the EV charging station sitting and sizing into a mix-integer linear programming problem. The proposed objective function is ameliorated through the proposed flowchart. The optimization function successfully introduces the costs of distribution expansion, EV station, voltage regulation as well as a well-designed protective device cost model to this problem. 

The problem sensitivity is not compromised after the convexification. The idea on the convex preservation of constraints always guarantees a global minimum in different test cases. Meanwhile, the computational time is greatly decreased with convex preservation. Through the numerical results, we realize that the voltage regulation cost is trying to favor the EV charging station placement at the end of branches. However, the protective device upgrade will cost less if more EV charging stations are installed at the main line of the feeder, trying to avoid branch ends. Numerical results also show that the protective device cost is not negligible in the total planning cost. At the end of numerical results, the proposed method illustrates that it is a flexible and versatile solution for the EV charging station placement no matter in developed countries or developing countries.  


\ifCLASSOPTIONcaptionsoff
  \newpage
\fi

\appendices
\section{Protective device costs}
\label{sec:protectioncostdistributionsystem}

The protective device costs are shown in Table \ref{tab:protinstmain} and \ref{tab:protacq} \cite{ref:meneses2013improving}. 

\begin{table}[!hbt]
	\renewcommand{\arraystretch}{1.3}

	\caption{Protective device installation and maintenance costs.}
	\label{tab:protinstmain}
	\centering
	\begin{tabular}{>{\centering\arraybackslash}m{2.5cm}>{\centering\arraybackslash}m{2cm}>{\centering\arraybackslash}m{2.5cm}}
		\hline
		\hline
		Device type & Install/uninstall cost (\$) & Annual maintenance cost  (\$)\\
		\hline
		Fuse& $1,000$ & $50$\\
		Recloser& $5,000$ & $2,500$\\
		Overcurrent relay& $1,000$ & $500$\\
		DORSR& $1,500$ & $750$\\
		\hline
		\hline
	\end{tabular}
\end{table}

\begin{table}[!hbt]
	\renewcommand{\arraystretch}{1.3}

	\caption{Protective device acquisition costs.}
	\label{tab:protacq}
	\centering
	\begin{tabular}{>{\centering\arraybackslash}m{2.5cm}>{\centering\arraybackslash}m{2cm}>{\centering\arraybackslash}m{1cm}}
		\hline
		\hline
		Device type & Current (A) & Cost (\$)\\
		\hline
		\multirow{5}{*}{Fuse}& $0\sim20$ & $400$\\
		& $21\sim50$ & $700$\\
		& $51\sim80$ & $850$\\
		& $81\sim100$ & $1,000$\\
		& $101\sim200$ & $1,100$\\
		\hline
		\multirow{5}{*}{Recloser}& $0\sim50$ & $15,000$\\
		& $51\sim100$ & $19,000$\\
		& $101\sim300$ & $22,000$\\
		& $301\sim500$ & $27,000$\\
		& $501\sim1,000$ & $30,000$\\
		\hline
		\multirow{5}{*}{Overcurrent relay}& $0\sim50$ & $4,000$\\
		& $51\sim100$ & $4,500$\\
		& $101\sim300$ & $5,000$\\
		& $301\sim500$ & $5,500$\\
		& $501\sim1,000$ & $6,000$\\
		\hline
		\multirow{5}{*}{DORSR}& $0\sim50$ & $6,000$\\
		& $51\sim100$ & $6,500$\\
		& $101\sim200$ & $7,000$\\
		& $201\sim500$ & $7,500$\\
		& $501\sim1,000$ & $8,000$\\
		\hline
		\hline
	\end{tabular}
\end{table}

\section{Derivation of the AC power flow linearization}
\label{sec:acpf_linear}

Nodal currents can be expressed by the admittance matrix and nodal voltages:

\begin{equation}
\begin{pmatrix} 
  I_S\\ 
  I_N
\end{pmatrix}
=\begin{pmatrix} 
  Y_{SS}     & Y_{SN}\\ 
  Y_{NS} & Y_{NN}
\end{pmatrix}
\cdot \begin{pmatrix} 
  V_S\\ 
  V_N 
\end{pmatrix}
\end{equation}

where $S$ represents the slack node and $N$ is the set of remaining nodes. Each nodal current is related to the voltage by the following ZIP model:

\begin{equation}
I_k=\frac{S_{Pk}^*}{V_k^*}+h \cdot S_{Ik}^*+h^2 \cdot S_{Zk}^* \cdot V_k
\end{equation}

We linearize the AC power flow equation and express the voltage as a function of the power injected in a closed rectangular form \cite{ref:garces2016linear}:

\begin{equation}
A+B \cdot V_N^*+C \cdot V_N=0
\label{eq:linearizeACpf}
\end{equation}

\noindent with $A=Y_{NS} \cdot V_S=2h \cdot S_{PN}^*-h \cdot S_{IN}^*$, $B=h^2 \cdot \text{diag}(S_{PN}^*)$, $C=Y_{NN}-h^2 \cdot \text{diag} (S_{ZN}^*)$, where $V_N$ is the vector of non-slack bus voltages, $S_{ZN}$, $S_{IN}$ and $S_{PN}$ are the complex power injection of constant impedance load, constant current load and constant power load at non-slack buses, $h=1/V_{nom}$.

\bibliographystyle{IEEEtran}
\bibliography{IEEEabrv,EVstation}

\vspace{-10 mm}
\begin{IEEEbiography}[{\includegraphics[width=1.2in,height=1.35in,clip,keepaspectratio]{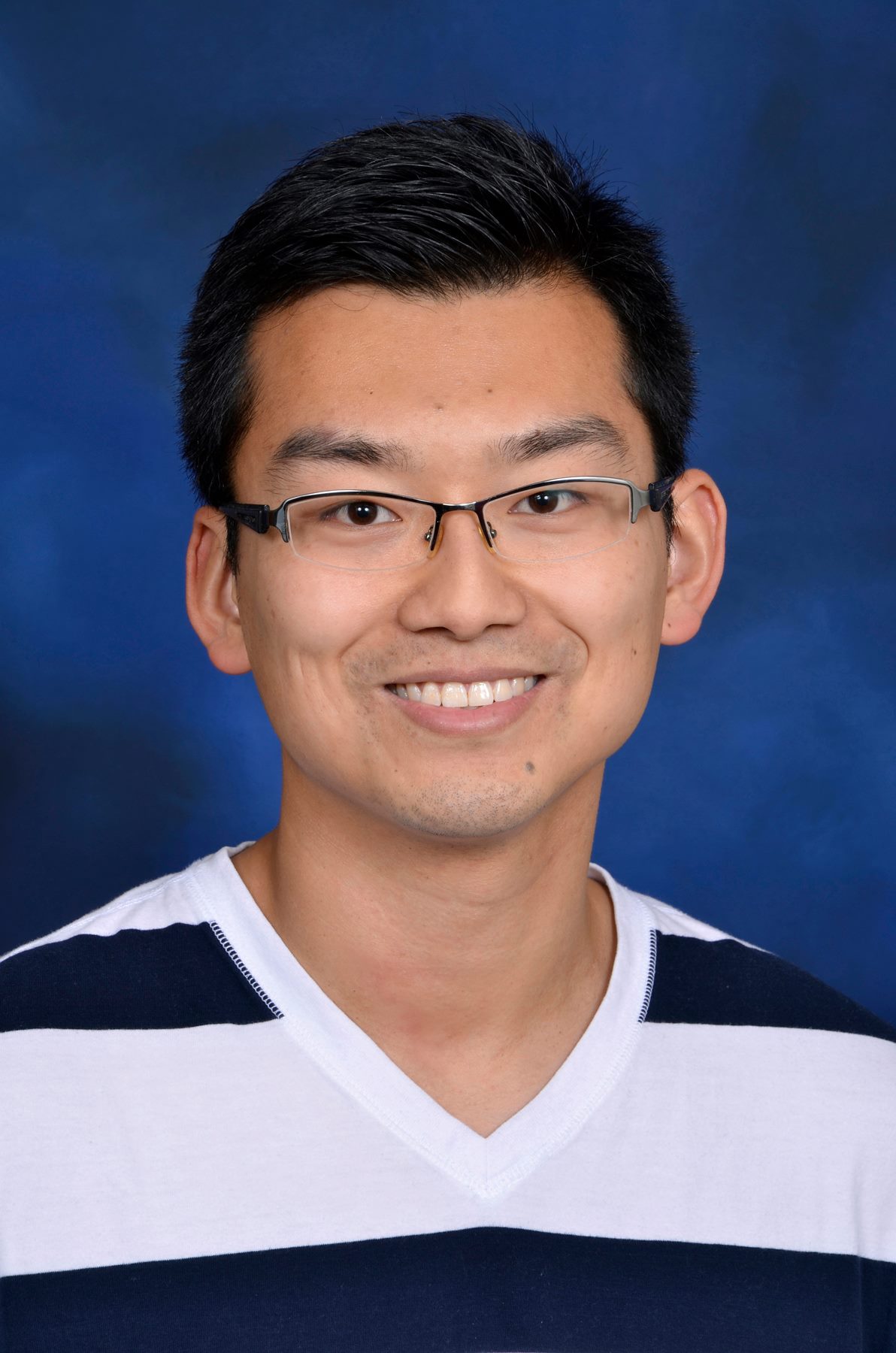}}]{Qiushi Cui} (S'10-M'18) received the M.Sc. degree from Illinois Institute of Technology in 2012, and the Ph.D. degree from McGill University in 2018, both in Electric Engineering. He was a research engineer at OPAL-RT Technologies Inc. since Nov. 2015 and sponsored by Canada MITACS Accelerate Research Program at the same company until Nov. 2017. Since Jan. 2018, he joined Arizona State University as a postdoctoral researcher.  

His research interests are in the areas of big data applications in power system protection, power system modeling, microgrid, EV charging station placement, and real-time simulation in power systems. 

\end{IEEEbiography}
\begin{IEEEbiography}[{\includegraphics[width=1in,height=1.25in,clip,keepaspectratio]{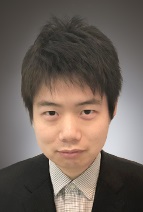}}]{Yang Weng} (M'14) received the B.E. degree in electrical engineering from Huazhong University of Science and Technology, Wuhan, China; the M.Sc. degree in statistics from the University of Illinois at Chicago, Chicago, IL, USA; and the M.Sc. degree in machine learning of computer science and M.E. and Ph.D. degrees in electrical and computer engineering from Carnegie Mellon University (CMU), Pittsburgh, PA, USA.

After finishing his Ph.D., he joined Stanford University, Stanford, CA, USA, as the TomKat Fellow for Sustainable Energy. He is currently an Assistant Professor of electrical, computer and energy engineering at Arizona State University (ASU), Tempe, AZ, USA. His research interest is in the interdisciplinary area of power systems, machine learning, and renewable integration.

Dr. Weng received the CMU Dean’s Graduate Fellowship in 2010, the Best Paper Award at the International Conference on Smart Grid Communication (SGC) in 2012, the first ranking paper of SGC in 2013, Best Papers at the Power and Energy Society General Meeting in 2014, ABB fellowship in 2014, and Golden Best Paper Award at the International Conference on Probabilistic Methods Applied to Power Systems in 2016.

\end{IEEEbiography}
\vspace{-180 mm}
\begin{IEEEbiography}[{\includegraphics[width=1.1in,height=2.25in,clip,keepaspectratio]{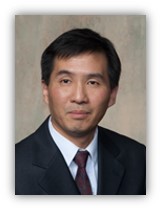}}]{Chin-Woo Tan} received the B.S. and Ph.D. degrees in electrical engineering, and the M.A. degree in mathematics from the University of California, Berkeley, CA, USA. 

Currently, he is Director of Stanford Smart Grid Lab. He has research and management experience in a wide range of engineering applications intelligent sensing systems, including electric power systems, automated vehicles, intelligent transportation, and supply chain management. His current research focuses on developing data-driven methodologies for analyzing energy consumption behavior and seeking ways to more efficiently manage consumption and integrate distributed energy resources into grid. Dr. Tan was a Technical Lead for the LADWP Smart Grid Regional Demonstration Project, and a Project Manager with the PATH Program at UC Berkeley for 10 years, working on intelligent transportation systems. Also, he was an Associate Professor with the Electrical Engineering Department at California Baptist University. 

\end{IEEEbiography}

\end{document}